\def\aa{{A\&A}}
\def\aas{{A\&AS}}
\def\aj{{AJ}}
\def\annrev{{ARA\&A}}
\def\apj{{ApJ}}
\def\apjs{{ApJS}}
\def\baas{{BAAS}}
\def\mnras{{MNRAS}}
\def\nat{{Nature}}
\def\pasp{{PASP}}

\documentclass[cup5b]{caps}
\usepackage{graphicx}
\usepackage{amssymb}
\usepackage{ociwsymp4e}  

\HeadText{Jeffrey L. Linsky}  

\begin{document}

\pagenumbering{arabic}

\author[]{JEFFREY L. LINSKY\\JILA/University of Colorado and NIST, Boulder, CO, 
USA} 

%
%

\chapter{D/H as a Measure of Chemical Inhomogeneity in our Galaxy}

\begin{abstract}

Accurate measurements of interstellar deuterium abundances along lines of sight
extending out to several hundred parsecs by FUSE and other instruments is
making D/H a useful tool for understanding Galactic chemical evolution. We find
that the gas inside of the Local Bubble is chemically homogeneous, but that
large variations in D/H beyond the Local Bubble are real and challenge present
concepts of chemical evolution. A new set of models is needed that will include
spatially dependent infall of relatively unprocessed material, depletion of D 
onto grains, and appropriate mixing timescales.

\end{abstract}

\section{Tools of the trade}

Since accurate measurements of the deuterium abundances are important for
understanding the early history of the Universe and Galactic chemical
evolution, investigators have pursued a variety of techniques to measure D/H in
different locations. As reviewed by Linsky (2003), these techniques include the
study of (1) deuterated molecules in cold interstellar clouds, (2) HD/H$_2$ in
the ISM, (3) deuterium Balmer-$\alpha$ emission in the Orion Nebula, (4) 92 cm
emission from the hyperfine transition line of D~I in the ISM, (5) D/H in the
solar photosphere, (6) D/H in comets, meteorites, and planetary atmospheres,
and (7) deuterium Lyman line absorption in warm interstellar gas. The last
technique has proven to be the most accurate because in the warm ($T\sim
7,000$~K) component of the ISM there are very few molecules, depletion of H and
D onto grains is minimal, and the ionization equilibria of D and H are very
similar. Thus measurements of the D~I and H~I column densities, N(D~I) and
N(H~I), directly provide the D/H total abundance ratio. However, the factor of
$10^5$ difference in the abundances of D and H means that the interstellar H
Lyman lines are all very optically thick, whereas the D Lyman lines are
optically thin. Thus this technique must be applied with care as the extremely
thick Lyman lines of H may contain unresolved absorption components not seen in
lines of any other element and small errors in the assumed background continuum
can result in large errors in the H column density and thus the D/H ratio. 

Analysis of Lyman line absorption along many lines of sight toward different
stars have demonstrated the following requirements for accurate D/H
measurements: 

\begin{description}
\item[High S/N:] Signal/noise $\geq 30$ is highly desirable especially for
column densities N(H~I) $< 10^{19}$ cm$^{-2}$ when the wings of the Ly-$\alpha$
line are not yet strong and N(H~I) must be measured from the outer edges of the
absorption line core. 

\item[Knowledge of the line of sight:] The low mass H and D lines are broadened
primarily by thermal motions and thus provide little information on the
velocity structure along the line of sight. The intrinsically narrow lines of
heavier ions like Mg~II and Fe~II provide information on the Doppler motions
(systematic and turbulent) along the line of sight. It is important to include
this information to the analysis of the very saturated H lines since small
errors in the nonthermal broadening of these lines due to not including the
Doppler shifts of H~I along the line of sight can translate into very large
errors in N(H~I). 

\item[High spectral resolution:] To disentangle the velocity structure along a
line of sight, spectral resolution comparable to the spacing of the velocity
components is required. Welty et al. (1996) showed that even resolutions of
0.3--1.0 km~s$^{-1}$, are inadequate to fully resolve the interstellar velocity
structure observed with the Ca~II K line. It is likely that the velocity
structure for warm interstellar H~I gas will have similar velocity structure.
Since the thermal width of the H~I lines is about 10 km~s$^{-1}$, the inferred
D/H ratio will be an average value for the line of sight. It is essential,
however, to include the velocity structure when deriving N(H~I) from the shape
of the interstellar absorption. A resolution better than 5 km~s$^{-1}$ is
needed to measure N(H~I) from the shape of the outer edges of the interstellar
absorption. 

\item[Low N(H~I):] Since the H and D Lyman lines are separated by only 82
km~s$^{-1}$, there will be a ``horizon'' for each Lyman line set by the value
of N(H~I) when the interstellar H absorption becomes so broad that the D line
cannot be observed. The horizon is $5\times 10^{18}$ cm$^{-2}$ for Ly-$\alpha$
and $3\times 10^{19}$ cm$^{-2}$ for Ly-$\beta$. While HST can observe the
Ly-$\alpha$ line, the Far Ultraviolet Spectrometer Explorer (FUSE) is needed to
study Ly-$\beta$ and the higher Lyman lines which have more distant horizons.
Depending on the value of N(H~I) along the line of sight and how accurately it
can be measured, various strategies have been employed to measure D/H bny
measuring N(D~I) from different lines than are used for measuring N(H~I) and by
using proxies like N(O~I) for estimating N(H~I) (see below). Also, N(H~I) can
be measured from the Lyman continuum absoprtion observed, for example, by EUVE.

\item[Knowledge of the ``continuum'':] For hot white dwarfs, subdwarfs, and OBA
stars, the stellar Lyman lines are absorption features, while for late-type
stars the Lyman lines are in emission. Since the interstellar absorption
is broad, one cannot simply interpolate the intrinsic stellar flux from the
emission outside of the Lyman line cores, but instead one must compute
the stellar line shape using a model atmosphere program or estimate the line
shape from comparison with other stars line the Sun. The former approach
typically used for the hotter stars requires that one know the stellar
parameters accurately, while the latter approach typically used for late-type
stars requires that one know the intrinsic shape of the line in the comparison
star. Both approaches are sources of systematic error in the inferred D/H
ratio. 

\item[Knowledge of the heliosphere and astrosphere:] The interaction via charge
exchange between the ionized solar (or stellar) wind and the inflowing
partially ionized ISM gas leads to additional neutral H absorption. The
heliospheric absorption is redshifted relative to the ISM for all viewing
angles and the astrospheric absorption is blueshifted (cf. Wood et al. 2002a).
Since the heliospheric and astrospheric absorption in H~I is not very optically
thick, there is negligible absorption in D and metal lines. The heliospheric
and astrospheric absorption can be very important because it broadens the total
absorption in the H Lyman lines. If not properly modelled, this extra
absorption would appear to be interstellar, raising the inferred N(H~I) and
thereby lowering the inferred D/H ratio for the line of sight. 

\item[The problem is H:] For most Galactic lines of sight studied so far, the
errors in N(H~I) usually exceed those in N(D~I), since the H lines are very
optically thick. Also, the hard to estimate systematic errors usually exceed
random measurement errors. One should keep this in mind when using D/H
measurements reported in the literature. 

\end{description} 

Table 1.1 lists the various satellites and instruments that have been used to 
measure D/H using UV and far-UV spectrographs. The fourth column of the table 
lists the target stars observed giving their spectral types and whether they 
are a white dwarf (WD) or subdwarf (sd). D/H values are published for the ISM
along these lines of sight, but for some of the older data sets (especially
IUE) the D/H values are not very accurate and will not be used later in this
paper. 

 \begin{table}
  \caption{Moderate and high resolution UV spectroscopic instruments.}
    \begin{tabular}{lcccc}
     \hline \hline
{Instrument} & {Spectral}    & {Resolution}    & {D/H Target}   & {Observing} \\
             & {Range (\AA)} & {(km s$^{-1}$)} & {Stars}        & {Dates} \\ 
\hline\hline
Copernicus & ~~900--3150   & 15            & 11 OB, 6 FGK & 1972--1981 \\
IUE/SWP-HI & 1170--1950   & 25            & 2 GK         & 1978--1996 \\
HST/GHRS   & 1170--3200   & ~3.5          & 15 FGK, 1 A, 1 WD& 1990--1997 \\
ORFEUS     & ~~912--1410   & 30            & 1 sdO        & 1993, 1996 \\
IMAPS      & ~~930--1150   & ~4.0          & 3 O          & 1993, 1996 \\
HST/STIS   & 1170--3200   & ~3.0          & 1 K, 1 B, 3 WDs& 1997--2010? \\
FUSE       & ~~912--1180   & 20            & 8 WDs, 5 OB  & 1998--?  \\ 
HST/COS    & 1150--3200   & 15            &              & 2005? \\
     \hline \hline
    \end{tabular}
  \label{instrument-table}
 \end{table}


\section{D/H, D/O, and D/O inside of the Local Bubble}

The Sun is located in a portion of the Galactic disk known as the Local Bubble
(LB). The LB extends for 50--200 pc in different directions with its outer edge
defined by a ``wall'' of relatively cold material seen as NaI absorption (Sfeir
et al. 1999). Most of the material inside of the LB is very low density hot gas
($\log T \approx 6.0$) that is likely produced by the winds of hot stars and
the ejecta of supernovae in the Sco-Cen Association. The interstellar gas flows
past the Sun at 28.1 km~s$^{-1}$ from the direction $l=12.4^{\circ}$,
$b=+11.6^{\circ}$ (Frisch et al. 2002), which is close to the center of the
Sco-Cen Association. 

Located inside of the LB are a number of warm ($T\sim 7,000$~K) partially
ionized clouds with total densities about 0.2 cm$^{-3}$. One such cloud is the
Local Interstellar Cloud (LIC) first identified by line of sight interstellar
velocities toward nearby stars that are consistent with a single vector
(Lallement \& Bertin 1992). The Sun is located just inside of the LIC, because
the flow of neutral He inside the heliosphere is consistent with the LIC flow
vector and no absorption at the predicted LIC velocity is seen toward stars in
the Galactic Center direction. The LIC is roughly spherical in shape with
dimensions 4.7 x 6.8 pc, $N_{HI} = 2.1\times 10^{18}$ cm$^{-2}$, and $n_{HI}
\approx 0.10$ cm$^{-3}$ (Redfield \& Linsky 2000). 

Table 1.2 summarizes the D/H measurements towards stars located inside of the
LB. Since the photoionization continua of H and D overlap, the selfshielding of
D~I and H~I to far ultraviolet radiation is the same, their ionization
equilibria are the same, and the ratio of column densities, N(D~I)/N(H~I), will
give the number density ratio DI/HI and thus D/H. The first column in the
table lists the measured values of DI/HI in parts per million (ppm). The HST
results were obtained from the analysis of the Lyman-$\alpha$ line profiles
observed towards G and K stars mostly with the GHRS. The DI/HI ratios in 
Linsky (1998) are the weighted means for lines of sight to 12 stars with
velocity components consistent with the LIC velocity vector, and for lines of
sight to 17 stars with velocity components for many clouds located inside of
the LIC. The listed uncertainties are for the means ratios and do not give the
variance in the individual lines of sight. Also listed are DI/HI values for
lines of sight toward 6 white dwarfs located inside of the LIC that have been
observed by FUSE. These DI/HI ratios are obtained from analyses of the higher
members of the Lyman series, although in several cases the N(H~I) values are
from the analysis of HST or EUVE spectra.

 \begin{table}
  \caption{Measurements of D/H inside the Local Bubble.}
    \begin{tabular}{lcccl}
     \hline \hline
{DI/HI}      & {Instrument} & {Targets}     & {Location} & {Reference} \\
{(ppm)}      & {Used}       &               &            & \\ \hline
$15\pm 1$    & HST/GHRS     & 11 GK, 1 WD   & inside LIC & Linsky (1998)\\
$14.7\pm 1.8$& HST/GHRS     & 15 GK, 2 WD   & $< 100$ pc & Linsky (1998)\\
$16.0\pm 2.5$& FUSE         & WD1634-573 (DO)& 37 pc     & Wood et al. (2002b)\\
15.1:        & FUSE         & WD2211-495 (DA)& 53 pc & H\'ebrard et al. (2002)\\
$16.6\pm 1.4$& FUSE         & HZ43A (DA)    & 68 pc      & Kruk et al. (2002)\\
$16.6^{+4.5}_{-3.0}$& FUSE  & G191-B2B (DA1)& 69 pc     & Lemoine et al. (2002)\\
14.1:        & FUSE         & WD0621-376 (DA)& 78 pc     & Lehner et al. (2003)\\
$15.1^{+3.9}_{-3.3}$&FUSE   & GD 246 (WD)   & 79 pc   & Oliveira et al. (2003)\\ 
     \hline \hline
    \end{tabular}
  \label{LBDH-table}
 \end{table}

Table 1.3 summarizes the measured OI/HI and DI/OI ratios for lines of sight
entirely within the LB. As the most abundant element after hydrogen and helium,
oxygen is a good measure of the total metal abundance. Since the ionization
potentials of neutral oxygen and hydrogen are nearly the same and the two
neutrals are closely coupled by change exchange reactions, the column density
ratio, N(O~I)/N(H~I), should accurately measure the OI/HI ratio in the gas
phase. Some oxygen in warm clouds will be locked up in grains, however, and the
total O/H ratio will therefore be somewhat larger than OI/HI in the gas phase.
The N(D~I)/N(O~I) ratio, equaling DI/OI in the gas phase, is especially useful
because the column densities of D~I and O~I differ by only a factor of 30,
rather than a factor of $10^5$ in the case of DI/HI, so that one can measure
optically thin D~I and O~I lines in FUSE spectra. Table 1.3 lists measured line
ratios in FUSE spectra of seven white dwarfs located within the LB. Also listed
in the table are ratios for the Capella line of sight and mean values for 33
lines of sight inside of the LB measured from HST spectra (GHRS or STIS). The
HST ratios have larger errors bars than the FUSE ratios, because N(O~I) is
measured using the optically thick 1302~\AA\ line. The recently measured solar
value of OI/HI by Allende Prieto et al. (2001) is also given in the table. 

 \begin{table}
  \caption{Measurements of OI/HI and DI/OI inside the Local Bubble.}
    \begin{tabular}{lccccl}
     \hline \hline
{OI/HI}    &  {DI/OI}    & {Instr.}& {Targets}   & {Location}  & {Reference} \\
{($10^{-4}$)}& {($10^{-2}$)}& {Used} &           &             & \\ 
\hline
$4.6\pm 0.7$& $3.5\pm 0.3$    &FUSE& WD1634-573 (DO)& 37 pc&Wood et al. 
(2002b)\\
3.8::      & $4.0\pm 0.6$     &FUSE& WD2211-495 (DA)& 53 pc& 
H\'ebrard et al. (2002)\\
$3.6\pm 0.4$& $4.6\pm 0.5$    &FUSE& HZ43A (DA)     & 68 pc& Kruk et al. 
(2002)\\
$4.6\pm 0.7$& $3.5\pm 0.4$ &FUSE& G191-B2B (DA1) & 69 pc& Lemoine et al. 
(2002)\\
3.6:       & $3.9^{+0.6}_{-0.5}$ &FUSE& WD0621-376 (DA)& 78 pc& Lehner et al. 
(2003)\\
$3.63^{+0.77}_{-0.67}$& $4.17^{+1.2}_{-1.0}$ &FUSE& GD 246 (WD)& 79 pc & 
Oliveira et al. (2003)\\ 
     & $5.13^{+2.20}_{-1.69}$& FUSE& WD2331-475 (DA)& 82 pc & Oliveira et al. 
(2003)\\
           &             &        &              &    &       \\
$6.0^{+6.1}_{-3.0}$&$2.65^{+2.7}_{-1.3}$ &HST & Capella (GIII) & 12 pc& 
Wood et al. (2002c)\\ 
$2.5^{+1.9}_{-1.1}$& $6.0^{+4.7}_{-2.6}$& HST & 33 stars & $<100$ pc & Redfield 
(2003)\\
$4.90\pm 0.56$&          & Opt           & Sun     &  &Allende Prieto et al. 
(2001)\\
     \hline \hline
    \end{tabular}
  \label{oxygenLB-table}
 \end{table}

Consideration of these results lead me to the following conclusions:

\begin{enumerate}
\item There are no significant variations in DI/HI observed for lines of
sight inside of the LB.

\item We adopt the value D/H = $15\pm 2$ ppm as representative for the LB.

\item There is no evidence for significant variations in OI/HI or DI/OI inside 
of the LB. The scatter is consistent with measurement errors. 

\item Inside the LB OI/HI = $(4.0\pm 0.4)\times 10^{-4}$  
and DI/OI = $(4.1\pm 0.4)\times 10^{-2}$. 

\item DI/OI is more accurately measured than OI/HI.

\end{enumerate}

\section{D/H, D/O, and O/H outside of the Local Bubble}

As one proceeds beyond the LB, a very different picture emerges. Table 1.4
summarizes the published measurements of DI/HI obtained from analyses of far
ultraviolet spectra obtained with the FUSE and Copernicus satellites and with
IMAPS and ORFEUS for lines of sight to more distant white dwarfs, OB stars, a
Wolf-Rayet star and a hot subdwarf. Although the cited errors are larger than
for the sightlines within the LB, the range of DI/HI ratios is very large with
some values well below the LB ratio and some above. All of the measured DI/HI
ratios lie below the mean values for lines of sight to four or five QSOs cited
by O'Meara et al. (2001) and very recently by Kirkman et al. (2003). Figure 1
plots the D/H values obtained for sightlines inside and outside of the LB and
toward QSOs. 

 \begin{table}
  \caption{Measurements of DI/HI beyond the Local Bubble.}
    \begin{tabular}{lcccl}
     \hline \hline
{DI/HI}    & {Instrument}& {Targets}     & {Location}   & {Reference} \\
{(ppm)}    & {Used}      &               &              &           \\ 
\hline
$13.9\pm 1.0$& FUSE      & BD+28$^{\circ}$4211 (sdO)& 104 pc  & Sonneborn et 
al. (2002)\\
$5.0\pm 1.6$&Copernicus  & $\theta$ Car (B0V)& 135 pc   & Allen et al. (1992)\\
$21.4\pm 4.1$& FUSE      & Feige 110 (WD)& 180 pc      & Friedman et al. 
(2002)\\
$13\pm 2.5$& Copernicus  & $\gamma$ Cas (B0IV) & 188 pc & Ferlet et al. (1980)\\
$7.6\pm 2.5$& Copernicus & $\lambda$ Sco (B2IV)& 216 pc & York et al. (1983)\\
$21.8\pm 2.0$& IMAPS     & $\gamma^2$ Vel (WC)& 258 pc  & Sonneborn et al. 
(2000)\\
$12\pm 4$   & ORFEUS     & BD +39 3226   & 270 pc       & Bluhm et al. (2000)\\
$14.2\pm 1.5$& IMAPS     & $\zeta$ Pup (O5I) & 430 pc   & Sonneborn et al. 
(2000)\\
$7.4^{+1.2}_{-0.9}$&IMAPS& $\delta$ Ori A (O9II)& 500 pc& Jenkins et al. 
(1999)\\
$8.5^{+3.4}_{-2.4}$&FUSE & HD 195965 (B1Ib)& 800 pc     & Hoopes et al. (2003)\\
$7.8^{+5.2}_{-2.5}$&FUSE & HD 191877 (B0V)& 2.2 kpc     & Hoopes et al. (2003)\\
           &             &               &              & \\
$30\pm 4$  & Keck        & 4 QSOs        &              & O'Meara et al. 
(2001)\\
$27.8^{+4.4}_{-3.8}$& Keck & 5 QSOs      &              &Kirkman et al. (2003)\\
$20\pm 4$  &             & Protosolar    & here         & Geiss et al. (2002)\\
     \hline \hline
    \end{tabular}
  \label{DHbeyondLB-table}
 \end{table}

Are the D/H values obtained for sightlines outside of the LB accurate, or are
systematic errors, perhaps associated with the very saturated H Lyman lines,
responsible for the large scatter in D/H. To address this question, I consider
the DI/OI ratio, which is generally considered to be a good proxy for D/H. Table
1.5 summarizes the published DI/OI ratios for lines of sight extending beyond
the LB obtained from the analysis of FUSE, Copernicus, and IMAPS data. These
results show a wide range of DI/OI values like the direct measurments of the
DI/HI ratio. Could the wide range in the DI/OI ratios be due to a wide range in
the OI/HI ratios? Meyer, Jura, and Cardelli (1998) measured OI/HI using HST
observations of the weak intersystem O~I 1356~\AA\ line for sightlines
extending out to 1.5 kpc. (See Moos et al. 2002 for a correction to the
oscillator strength of this line.) He found very little variation in OI/HI in
this sample of sightlines. The OI/HI ratios measured from FUSE, Copernicus, and
IMAPS data listed in Table 1.5 are more widely scattered, but are consistent
with the Meyer et al. (1998) mean ratio. 

 \begin{table}
  \caption{Measurements of OI/HI and DI/OI beyond the Local Bubble.}
    \begin{tabular}{lccccl}
     \hline \hline
{OI/HI}    &  {DI/OI}  & {D/H}    & {Targets}     & {Location}  & {Reference} \\
{($10^{-4}$)}& {($10^{-2}$)}& {Inst.} &          &              &           \\
\hline
$2.4\pm 0.3$& $5.9\pm 0.7$& FUSE& BD+28$^{\circ}$4211 & 104 pc  & 
Sonneborn et al. (2002)\\
         &$4.57^{+2.22}_{-1.63}$& FUSE& HZ21 & 115 pc & Oliveira et al. (2003)\\
    &$3.24^{+3.27}_{-2.06}$& FUSE& Lan23     & 122 pc & Oliveira et al. (2003)\\
$3.9\pm 0.8$& $5.5^{+0.8}_{-0.7}$& FUSE& Feige 110    & 180 pc & 
Friedman et al. (2002)\\
$3.67\pm 0.62$& $3.5\pm 0.9$& Coper& $\gamma$ Cas & 188 pc & Ferlet, Meyer \\
$2.82\pm 0.46$& $2.62\pm 0.71$& IMAPS& $\delta$ Ori A & 500 pc& 
Jenkins, Meyer \\
$6.61^{+1.03}_{-1.11}$& $1.29^{+0.51}_{-0.37}$& FUSE & HD 195965 & 
800 pc     & Hoopes et al. (2003)\\
$3.09^{+1.98}_{-0.98}$& $2.51^{+2.14}_{-0.86}$& FUSE & HD 191877 & 
2.2 kpc     & Hoopes et al. (2003)\\
 & & & & & \\
$3.43\pm 0.15$&          & HST           &13 OB stars&$<1.5$ kpc&Meyer et al. 
(1998)\\
$4.90\pm 0.56$&          & Opt           & Sun     &  &Allende Prieto et al. 
(2001)\\
     \hline \hline
    \end{tabular}
  \label{oxygenbeyondLB-table}
 \end{table}

These results suggest an emerging picture of D/O, O/H, and D/H in the Galaxy:
\newpage

\begin{itemize}

\item Unlike the observed constant value inside of the LB, the gas phase
measurements of DI/OI outside of the LB show very credible variations with a
range (1.3--5.9)$\times 10^{-2}$. This range is more than a factor of 4! 

\item By contrast, the HST gas phase observations of OI/HI beyond the LB show 
very little variation with a mean value $(3.43\pm 0.15)\times 10^{-4}$. 
Measurements of OI/HI based on data from
FUSE and other satellites show some scatter but are consistent with the HST 
result. 

\item If we adopt the HST value for OI/HI in the gas phase, then the DI/OI
ratios imply that DI/HI lies in the range 4.3--20.2 ppm. Since some of the
oxygen is condensed on to grains, the inferred D/H ratios will be somewhat
higher. Moos et al. (2002) estimate that 25\% of the oxygen in the LB could 
reside in grains, and Jenkins (2003) provides additional support for this 
estimate. If this rough estimate is representative of the region within 
1--2 kps of the Sun, then the inferred range in D/H rises to 5.5--26.0 ppm,
which is similar to the directly measured range in D/H of 5.0--21.8 
ppm (Table 1.4). Thus the two methods of determining D/H are consistent, 
implying that no large systematic errors are present.

\item The question of whether the Sun is metal rich compared with the ISM has
been raised many times. If we adopt the Meyer et al. (1998) value for OI/HI in
the ISM gas phase and the most recent (and lowest) value for the solar O/H
ratio (Allende Prieto et al. 2001), then the Sun is oxygen rich by the factor
490 ppm/343 ppm = 1.43 (0.16 dex). If, on the other hand, we assume that 25\%
of oxygen is condensed on to grains, then the ratio becomes 490/457 = 1.07
(0.03 dex). Thus the Sun either has the present interstellar abundance of
oxygen, or, if there is minimal depletion of oxygen in the ISM, the Sun is
slightly oxygen rich. 

\item The extent to which primordial deuterium has been burned by nuclear 
reactions in stars to form heavier elements is often called the astration
factor, A = (D/H)$_{primordial}$/(D/H)$_{ISM}$. If we assume that the most 
recent value of D/H for 5 quasars with low metal abundances is close to the 
primordial value, then in the LB we have 
A = (D/H)$_{primordial}$/(D/H)$_{LB}$ = $(27.8\pm 4.1)/(15\pm 2) = 
1.85\pm 0.37$.

\item Beyond the LB and extending out to 1--2 kpc, the astration factors appear 
to have a wide range of values from 1.3 to 5.6. This very large range in the astration 
factor over a small portion of our Galaxy challenges contemporary Galactic 
chemical evolution models as I will describe below.

\item Since deuterium is destroyed and oxygen created over time, one might 
expect that the O/H and D/H ratios are inversely correlated. In fact there 
is as yet little evidence of any correlation (Moos et al. 2002). Why? Could 
there be something missing in the Galactic chemical evolution models?

\end{itemize}

  \begin{figure}
    \centering
    \includegraphics[width=10cm,angle=0]{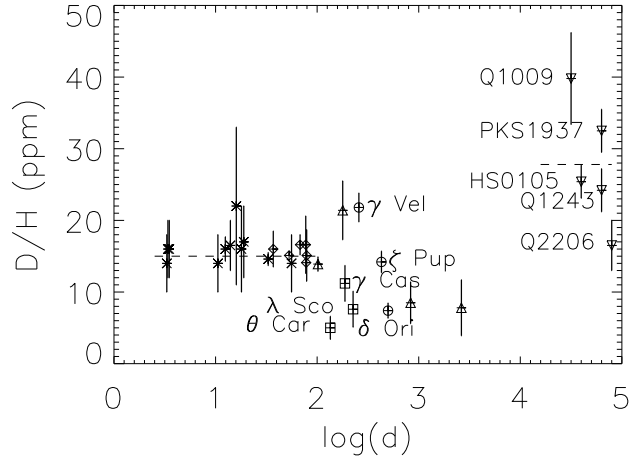}
    \caption{A summary of D/H measurements obtained with HST (asterisks), FUSE 
inside the LB (diamonds), FUSE outside of the LB (triangles), Copernicus 
(squares), and Keck (upside down triangles) with distance d in units of 
parsecs. The quasar lines of sight studied with Keck are not plotted at their 
correct distances. The dashed lines refer to the mean D/H ratios inside the LB 
and for the five QSO lines of sight.}
    \label{idl.ps}
  \end{figure}

\section{Can the D/O measurements be reconciled with Galactic chemical 
evolution?}

\subsection{Are the deuterium and oxygen abundances anticorrelated?}

If galactic chemical evolution occurs in a closed box in which the original gas
is nearly primordial and evolution proceeds without the input of extragalactic
material, then one expects D/H to decrease monotonically with time and O/H to
increase. In this simplest of chemical evolution models, one would expect to
see an anticorrelation of D/H and O/H when one samples different parcels of gas
with different chemical histories. This scenario is rather simple, but it does
suggest that we search for an anticorrelation of D and O in he FUSE data. Moos
et al. (2002) did this by plotting O/H vs D/H and D/O vs D/H for the nine
sightlines for which good measurements were available. They concluded that
there was no significant anticorrelation, although there is a hint of
decreasing D/O with increasing O/H. Steigman (2003) also says that there is a
hint of an anticorrelation between D/O and O/H. Observations of more sightlines
is needed to verify or refute this hint. 

\subsection{Galactic chemical evolution out of the box}

Geiss, Gloeckler, and Charbonnel (2002) have developed an empirical approach to 
explaining the present abundances of D/H, O/H, N/H, and $^3$He/H in the LIC 
that has interesting implications. They start with the abundances of the 
protosolar nebula measured in the most primitive material in the solar 
system, specifically D/H = $20\pm 4$ ppm and O/H = $545\pm 107$ ppm. They also 
assume that at the beginning of the Galaxy, D/H = $30\pm 4$ ppm and O/H = 0, 
and that the rate of Galactic chemical evolution is constant with time. Then
extrapolating the primordial abundances forward 4.6 Gyr to the present, they
predict that in the closed box approximation D/H = 12 ppm and O/H = 710 ppm.
Since D/H is too low, O/H too high, and the other ratios are also inconsistent
with the LIC values, they look for an alternative solution. The input of less
evolved gas from outside of the Galaxy would increase D/H and lower O/H,
leading to better agreement with the LIC values. Geiss et al. (2002) assumed
that the gas falling into the Galaxy in the region of the Sun during the last
4.6 Gyr was similar in composition to the LMC with D/H = 27 ppm and O/H = 160
ppm (30\% of the protosolar nebular value). The abundances of D/H, O/H, N/H,
and $^3$He/H are consistent with LIC values if the total gas includes a mixture
of 30--44\% infalling material with the chemical composition of the LMC. 

This purely empirical model demonstrates the need for infalling gas to explain 
the abundances of deuterium and oxygen in the local region of our Galaxy. The 
observations of large variations in D/O over distance scales of several hundred 
parsecs would suggest spatially and perhaps time dependent differences in the 
infall rate and long mixing timescales. 

Recent theoretical models of chemical evolution in the Galaxy provide different
scenarios for the input of relatively unprocessed material into the Galaxy. For
example, Chiappini, Renda, and Matteucci (2002) compute a "two-infall" model to
explain the halo and thin disk. Their model explains the D/H, $^3$He/H, and
$^4$He/H ratios near the Sun. They find that the value of D/H is very sensitive
to both the infall rate and the star formation rate. Their predicted astration
rate is $\approx 1.5$ in the solar neighborhood and increases gradually to
$\approx 5$ in the thin disk at 2 kpc from Galactic Center. These astration
rates cover the range of astration that we already see within a few hundred pc
of the Sun. This suggests that the spatial variations in the infall and stellar
formation rates are much larger than assumed in the Chiappini et al. (2002)
models, and that the time scales for mixing are very long. 

\subsection{Have we ignored something important?}

Before I complete this summary of deuterium as a tool for measuring chemical
inhomogeneity in our Galaxy, let us consider whether there could be a
systematic error in the way that we measure the D/H ratio. There are four
important questions to answer: 

\begin{enumerate}

\item Could the deuterium abundances be very inaccurate? I do not think so 
because the D/H ratios measured directly from the Lyman lines and indirectly 
through the DI/OI ratio are in substantial agreement. Also the D/H measurements 
obtained with HST and FUSE toward many different types of stars by different 
authors using different analysis software are in excellent agreement for 
sightlines inside of the LB bubble. There is no reason for expecting that a 
major systematic error should begin just as we consider sightlines outside of
the LB.

\item Could the deuterium absorption lines actually be blueshifted stellar 
hydrogen? In the Copernicus observations of OB stars, one occasionally sees
transient extra absorption near the location of the deuterium lines (-81 km 
s$^{-1}$ from the hydrogen line), presumably produced by transient absorption
in the stellar wind. The FUSE onservations, on the other hand, are mostly of 
white dwarfs and hot subdwarfs that do not show low velocity wind features.
Thus there is no reason to presume that the deuterium lines are not genuine.

\item Could a significant amount of deuterium be tied up in molecules? In the
FUSE observations of white dwarfs very few H$_2$ lines and no HD lines are
observed. In particular, Jenkins et al. (1999) detected no HD lines toward 
$\delta$~Ori~A, a line of sight with low D/H.
Thus N(H$_2$)/N(HI) $\ll 1$ and N(HD)/N(HI) is very much smaller. 

\item Could a significant amount of deuterium be tied up in grains as 
originally suggested by Jura (1982)? Even in the 
warm ISM, grains are cold and the binding energy of deuterium to carbon or 
metals on the surface of grains is slightly larger than for hydrogen. A serious 
calculation of this process is needed to address this question. Draine (2003) 
calculates that extreme D-enrichment of carbonaceous grains is possible in the 
absence of grain destruction processes.

\end{enumerate}

\section{Conclusions}

Accurate measurements of interstellar deuterium abundances along lines of sight
extending out to several hundred parsecs by FUSE and other instruments is
making D/H a useful tool for understanding Galactic chemical evolution. I
believe that the new D/H measurements directly from the Lyman line absorption
and indirectly from the D/O ratio are credible and for the most part may not
have systematic errors. We find that the gas inside the Local Bubble is
chemically homogeneous, but that large variations in D/H beyond the Local
Bubble are real and challenge present concepts of chemical evolution. A new set
of models is needed that will include spatially dependent infall of relatively
unprocessed material and the timescales for mixing should be investigated. 

\section{Acknowledgements}

I would like to thank the FUSE Science Team for providing the beautiful data 
upon which much of this paper is based. I also thank Seth Redfield and Brian 
Wood for helpful discussions on this topic. This work is supported by NASA 
through grants to the Johns Hopkins University and to the University of
Colorado.


\begin{thereferences}{}

\bibitem{} Allen, M.~M., Jenkins, E.~B., \& Snow, T.~P., 1992, \apjs, 83, 261

\bibitem{} Allende Prieto, C., Lambert, D.~L., \& Asplund, M., 2001, \apj, 556, 
L63

\bibitem{} Bluhm, H., Marggraf, O., de Boer, K.~S., Richter, P., \& 
Heber, U., 2000, \aa, 352, 287

\bibitem{} Chiappini, C., Renda, A., \& Matteucci, F., 2002, \aa, 395, 789

\bibitem{} Draine, B.~T., 2003, in {\em Carnegie Observatories Astrophysics 
Series, Vol. 4: Origin and Evolution of the Elements}, ed. A. McWilliam \& M. 
Rauch (Cambridge: Cambridge Univ. Press), 000

\bibitem{} Ferlet, R., York, D.~G., Vidal-Madjar, A., \& Laurent, C., 1980, 
\apj, 242, 578

\bibitem{} Friedman, S.~D. et al. 2002, \apjs, 140, 37 

\bibitem{} Frisch, P.~C., Grodnicki, L., \& Welty, D.~E., 2002, \apj, 574, 834

\bibitem{} Geiss, J., Gloeckler, G., \& Charbonnel, C., 2002 \apj, 578, 862

\bibitem{} H\'ebrard, G. et al. 2002, Planetary and Space Science, 50, 1169

\bibitem{} Hoopes, C.~G., Sembach, K.~R., H\'ebrard, G., Moos, H.~W., \& 
Knauth, D.~C., 2003, \apj, 586, 1094

\bibitem{} Jenkins, E.~B., 2003, in {\em Carnegie Observatories Astrophysics 
Series, Vol. 4: Origin and Evolution of the Elements}, ed. A. McWilliam \& M. 
Rauch (Cambridge: Cambridge Univ. Press), 000

\bibitem{} Jenkins, E.~B., Tripp, T.~M., Wozniak, P.~R., Sofia, U.~J., \& 
Sonneborn, G., 1999, \apj, 520, 182

\bibitem{} Jura, M., 1982, in {\em Advances in Ultraviolet Astronomy}, ed. Y. 
Kondo, (NASA CP-2238), 54 

\bibitem{} Kirkman, D., Tytler, D., Suzuki, N., O'Meara, J.~M., \& Lubin, D., 
2003, astro-ph/0302006

\bibitem{} Kruk, J.~W. et al., 2002, \apjs, 140, 19

\bibitem{} Lallement, R. \& Bertin, P., 1992, \aa, 266, 479\\

\bibitem{} Lehner, N., Jenkins, E.~B., Gry, C., Moos, H.~W., Chayer, P., \& 
Lacour, S., 2003, submitted to \apj

\bibitem{} Lemoine, M., et al., 2002, \apjs, 140, 67

\bibitem{} Linsky, J.~L., 1998, Space Science Reviews, 84, 285

\bibitem{} Linsky, J.~L., 2003, Space Science Reviews, in press

\bibitem{} Meyer, D.~M., Jura, M., \& Cardelli, J.~A., 1998, \apj, 493, 222

\bibitem{} Moos, H.~W., et al., 2002, \apjs, 140, 3

\bibitem{} Oliveira, C.~M., H\'ebrard, G., Howk, J.~C., Kruk, J.~W., Chayer, 
P., \& Moos, H.~W., 2003, \apj, 587, 2350

\bibitem{} O'Meara, J.~M., Tytler, D., Kirkman, D., Suzuki, N., Prochaska, 
J.~X., Lubin, D., Wolfe, A.~M., 2001, \apj, 552, 718

\bibitem{} Redfield, S. \& Linsky, J.~L., 2000, \apj, 534, 825

\bibitem{} Redfield, S. 2003, PhD thesis, University of Colorado

\bibitem{} Sfeir, D.~M., Lallement, R., Crifo, F., \& Welsh, B.~Y., 1999, \aa, 
346, 785

\bibitem{} Sonneborn, G., Tripp, T.~M., Ferlet, R., Jenkins, E.~B., Sofia, 
U.~J., Vidal-Madjar, A., and Wozniak, P.~R. 2000, \apj, 545, 277

\bibitem{} Sonneborn, G., et al., 2002, \apjs, 140, 51

\bibitem{} Steigman, G., 2003, \apj, 586, 1120

\bibitem{} Welty, D.~E., Morton, D.~C., \& Hobbs, L.~M., 1996, \apjs, 106, 533

\bibitem{} Wood, B.~E., M\"uller, H.~R., Zank, G.~P., Linsky, J.~L., 2002a,
\apj, 574, 412

\bibitem{} Wood, B.~E., Linsky, J.~L., H\'ebrard, G., Vidal-Madjar, A., 
Lemoine, M., Moos, H.~W., Sembach, K.~R., \& Jenkins, E.~B., 2002b, \apjs, 140, 
91

\bibitem{} Wood, B.~E., Redfield, S., Linsky, J.~L., \& Sahu, M.~S., 2002c,
\apj, 581, 1168

\bibitem{}York, D.~G., 1983, \apj, 264, 172




\end{thereferences}
\end{document}